\documentclass[twocolumn,showpacs,preprintnumbers,amsmath,floatfix]{revtex4}

\usepackage{graphicx}
\usepackage{dcolumn}
\usepackage{bm}

\newcommand{\gfive}{\gamma_{5}}             
\newcommand{\cC}{{\cal C}}                  
\newcommand{\cE}{{\cal E}}                  
\newcommand{\cR}{{\cal R}}                  
\newcommand{\cS}{{\cal S}}                  
\begin{document}

\title{Low-Dimensional Long-Range Topological Charge Structure in the QCD Vacuum}
\author{I.~Horv\'ath$^{a}$, S.J.~Dong$^{a}$, T.~Draper$^{a}$,
        F.X.~Lee$^{b,c}$, K.F.~Liu$^{a}$, N.~Mathur$^{a}$, 
        H.B.~Thacker$^{d}$ and J.B. Zhang$^{e}$}
\affiliation{
 $^{a}$Department of Physics, University of Kentucky, Lexington, KY 40506 \\
 \centerline{$^{b}$Center for Nuclear Studies and Department of Physics,
       George Washington University, Washington, DC 20052} \\
 $^{c}$Jefferson Lab, 12000 Jefferson Avenue, Newport News, VA 23606 \\
 $^{d}$Department of Physics, University of Virginia, Charlottesville, VA 22901 \\
 \centerline{$^{e}$CSSM and Department of Physics, 
       University of Adelaide, Adelaide, SA 5005, Australia}}
\date{September 15, 2003}


\begin{abstract}
While sign-coherent 4-dimensional structures cannot dominate topological charge 
fluctuations in the QCD vacuum at all scales due to reflection positivity, it is
possible that enhanced coherence exists over extended space-time regions of lower 
dimension. Using the overlap Dirac operator to calculate topological charge density, 
we present evidence for such structure in pure-glue SU(3) lattice gauge theory. 
It is found that a typical equilibrium configuration is dominated by two oppositely-charged 
sign-coherent {\em connected} structures ({\it ``sheets''}) covering about 80\% 
of space-time. Each sheet is built from elementary 3-d cubes connected through 2-d 
faces, and approximates a low-dimensional curved manifold (or possibly a fractal 
structure) embedded in the 4-d space. 
At the heart of the sheet is a {\em ``skeleton''} formed by about 18\% of the most
intense space-time points organized into a global {\em long-range} structure,
involving connected parts spreading over maximal possible distances. We find that
the skeleton is locally 1-dimensional and propose that its geometrical properties 
might be relevant for understanding the possible role of topological charge 
fluctuations in the physics of chiral symmetry breaking.   

\end{abstract}
\maketitle

One of the important goals in hadronic physics is a detailed understanding
of vacuum structure in QCD. Local patterns in topological charge fluctuations 
represent an important aspect of this structure. Indeed, phenomena such as large 
$\eta'$ mass, $\theta$-dependence, and possibly spontaneous chiral symmetry breaking 
(S$\chi$SB) are directly related to vacuum fluctuations of topological charge. While 
lattice gauge theory provides a framework for first-principles non-perturbative studies 
of topological charge fluctuations, the extraction of meaningful and unbiased 
structural information has long been a difficult problem. However, the recent development 
of fermionic methods has improved this situation considerably and it is no longer necessary 
to rely on subjective procedures to smooth out the rough short-distance behavior of gauge 
fields. One way to proceed is to study the structure of low-lying Dirac eigenmodes 
and to infer the properties of underlying topological charge fluctuations indirectly. 
This was advocated in~\cite{Inimodes}, developed in~\cite{Hor02A}, and used in many recent 
studies (see e.g.~\cite{Followup,Other}).

The advances in implementing lattice chiral symmetry 
(for reviews see e.g.~\cite{Chiral}) 
have also made it possible
to pursue a {\em direct} approach to this problem. Indeed, with any  
$\gfive$-Hermitian lattice Dirac operator $D$ satisfying the Ginsparg-Wilson relation
$\{D,\gamma_5\} = D \gamma_5 D$ (e.g. the overlap operator~\cite{Neu98A}), one can 
associate a topological charge density operator~\cite{Has98A} 
\begin{equation}
   q_x \,=\, -\mbox{\rm tr} \,\gamma_5 \, (1 - \frac{1}{2}D_{x,x})
   \label{eq:15}  
\end{equation}
The global charge associated with this lattice density is strictly stable with respect
to generic local variations of the gauge field~\cite{Stable}, thus fully respecting 
the topological nature of this quantity. In addition, the above $q_x$ is special in 
several ways. For example, the index theorem is exactly satisfied with respect to the 
associated Dirac operator~\cite{Has98A}, and its renormalization properties are 
analogous to those in the continuum~\cite{Giu01A}. An important virtue of $q_x$ is that
it can be naturally eigenmode-expanded. While $q_x$ contains all fluctuations
up to the lattice cutoff, one can use the eigenmode expansion up to scale 
$\Lambda$ to define an {\it effective density}~\cite{Hor02B}, 
$  q_x^{(\Lambda)} \,\equiv\, 
   -\sum_{|\lambda|\le\Lambda a} (1 - \frac{\lambda}{2})\, c^{\lambda}_x,$ 
where $c^\lambda_x = \psi^{\lambda \,\dagger}_x \gamma_5 \psi^{\lambda}_x$ is the local 
chirality of the mode with eigenvalue $\lambda$. The ultraviolet fluctuations
are naturally filtered out in $q_x^{(\Lambda)}$ ({\it eigenmode filtering}~\cite{Hor02A}), 
and this density can be used to study the structure of topological charge fluctuations
at arbitrary low-energy scale $\Lambda$. By studying the effective densities, it was 
demonstrated~\cite{Hor02B} that topological charge at low energy is not locally 
concentrated in sign-coherent unit-quantized lumps (e.g.\!\! instantons). This was first 
predicted by Witten~\cite{Wit79A} and emphasized recently in~\cite{Hor02A}, 
where the first lattice evidence was presented.

In this work, we initiate the investigation of possible non-trivial structure
(order) in the {\it full\/} topological charge density Eq.~(\ref{eq:15}) for 
typical configurations contributing to the pure glue QCD path integral. This
is a qualitatively new step with some aspects worth emphasizing. 
(a) We evaluate the local operator $q_x$ at every space-time point, thus producing 
gauge invariant ``configurations'' of topological charge density. No processing 
of the underlying gauge configurations is involved and no bias is imposed. 
(b) Assuming that the well-defined space-time structure in $q_x$ exists, it has 
to be viewed as {\em fundamental} (rather than low-energy) since, unlike the case
of effective densities $q_x^{(\Lambda)}$, the use of a local 
operator does not introduce a new scale apart from the lattice cutoff 
already in place. Consequently, such structure could have both short and 
long-distance manifestations. (c) A fundamental structure of this type has not 
been observed before in an unbiased setting. Indeed, using the conventional
operators, the resulting space-time distributions of topological charge 
can hardly be distinguished from complete disorder. The common interpretation 
of this is that the physically relevant fluctuations might be obscured by the 
structureless ultraviolet noise arising due to entropy considerations (the problem 
of ultraviolet dominance).

In what follows, we will present evidence for the existence of fundamental structure
in topological charge density, and demonstrate some of its basic geometric features.
Needless to say, we do this with the aim that some of these properties might help
to advance our understanding of the QCD vacuum, and hence of the basic features of 
hadronic physics. With regard to the problem of ultraviolet dominance, we were 
motivated by the fact that, apart from beautiful properties mentioned above, 
the operator $q_x$ of Eq.~(\ref{eq:15}) differs from standard lattice discretizations 
in a very qualitative manner. In particular, it is well-known that lattice chiral 
symmetry implies non-ultralocality of lattice Dirac kernel $D$, i.e. non-zero 
coupling among fermionic variables at 
arbitrarily large distances~\cite{Nultr}. Similarly, it is believed that any $D_{x y}$ 
involves gauge paths (from $x$ to $y$) running arbitrarily far from $x$ and $y$. 
As a result, $q_x$ will receive (small) contributions from arbitrarily extended gauge 
loops ({\it chiral smoothing}~\cite{Hor02A}\,) and is expected to be much less sensitive 
to the ultraviolet noise than typical ultralocal operators. We find that chiral 
smoothing indeed provides enough suppression of ultraviolet noise for 
the underlying coherent structure to be revealed.

As a guide to identifying this structure, we note the fact that 
$\langle\, q(x) q(0)\,\rangle \le 0 \;, |x|>0\,$, in the continuum~\cite{SeSt}.
This implies that the topological charge cannot be predominantly concentrated in 
4-dimensional sign-coherent structures of finite physical size~\cite{Hor02B}. However, 
one cannot conclude that the local behavior of topological charge density is strictly 
disordered! Indeed, the negativity of the correlator does not exclude enhanced 
sign-coherence present on {\it lower-dimensional} subsets of 4-d space-time. In that 
case the structure could respect the negativity of the correlator by appropriate 
embedding of subsets with alternating sign into space-time. Consequently, we focus 
our search on the level of sign-coherence, and how it changes when concentrating on 
space-time subregions containing strong fields. In what follows we will refer 
to sign-coherence in $q_x$ simply as coherence.

\begin{table}[b]
  \centering
  \begin{tabular}{cccccc}
  \hline
  \multicolumn{1}{c}{ensemble}   &
  \multicolumn{1}{c}{$\beta$}    &
  \multicolumn{1}{c}{$a$ [fm]}   &
  \multicolumn{1}{c}{$V$}        &
  \multicolumn{1}{c}{$V_p$ [fm$^4$]} &
  \multicolumn{1}{c}{configs}    \\
  \hline
  ${\cE}_1$ & 5.91  & 0.110 & $12^4$  & 3.0   & 8\\
  ${\cE}_2$ & 6.07  & 0.082 & $16^4$  & 3.0   & 2\\
\hline
\end{tabular}
\vspace*{-0.01in}
\caption{Ensembles of Wilson gauge configurations. The global topological charges are
         (3,0,-2,0,2,-1,-1,3) for $\cE_1$ and (2,1) for $\cE_2$.}
\vspace*{-0.18in}
\label{table:1} 
\end{table}

\begin{figure*}[htb!]
\hspace*{-0.1in}
\includegraphics[height=4.8cm]{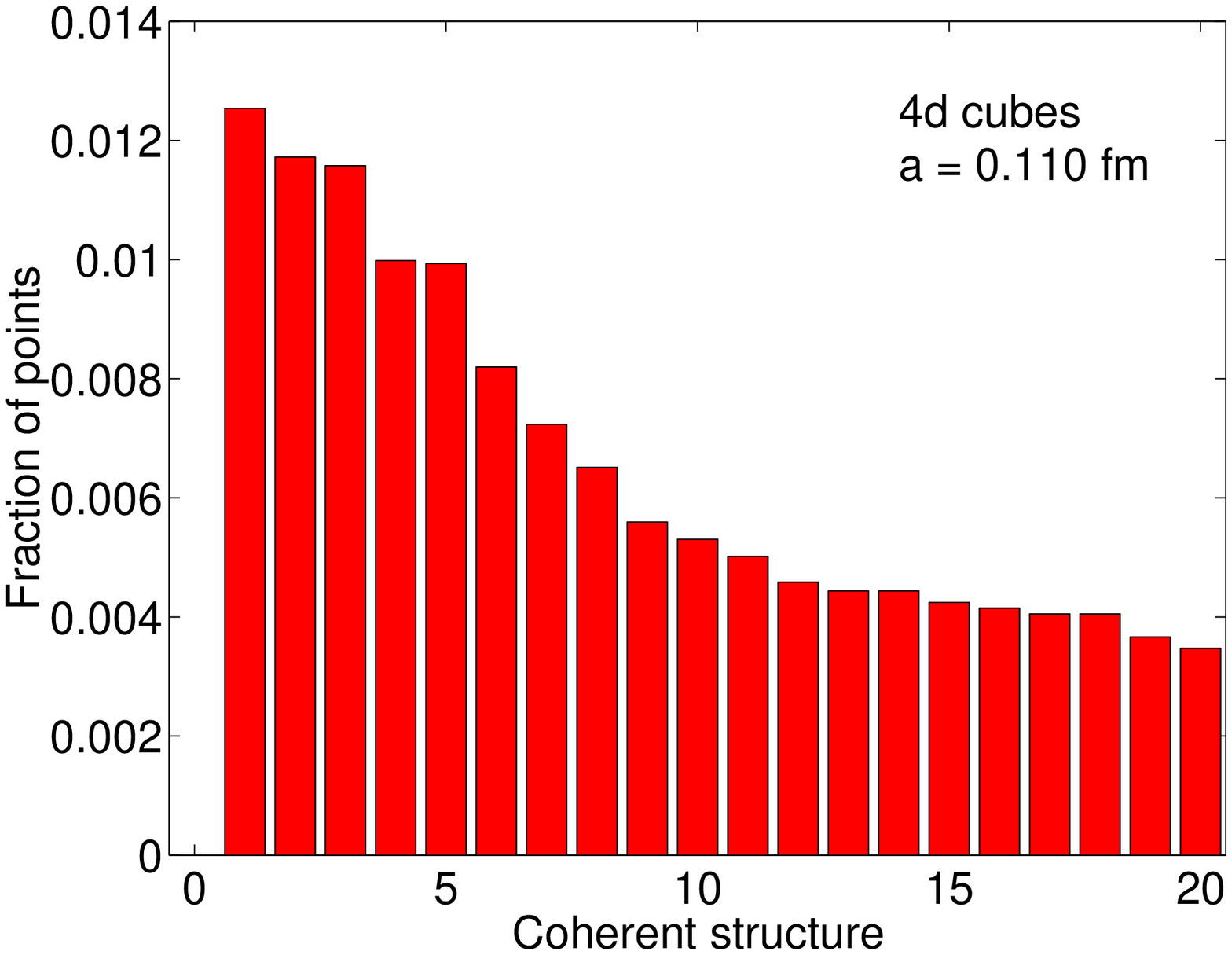}
\hspace*{-0.3in}
\includegraphics[height=4.8cm]{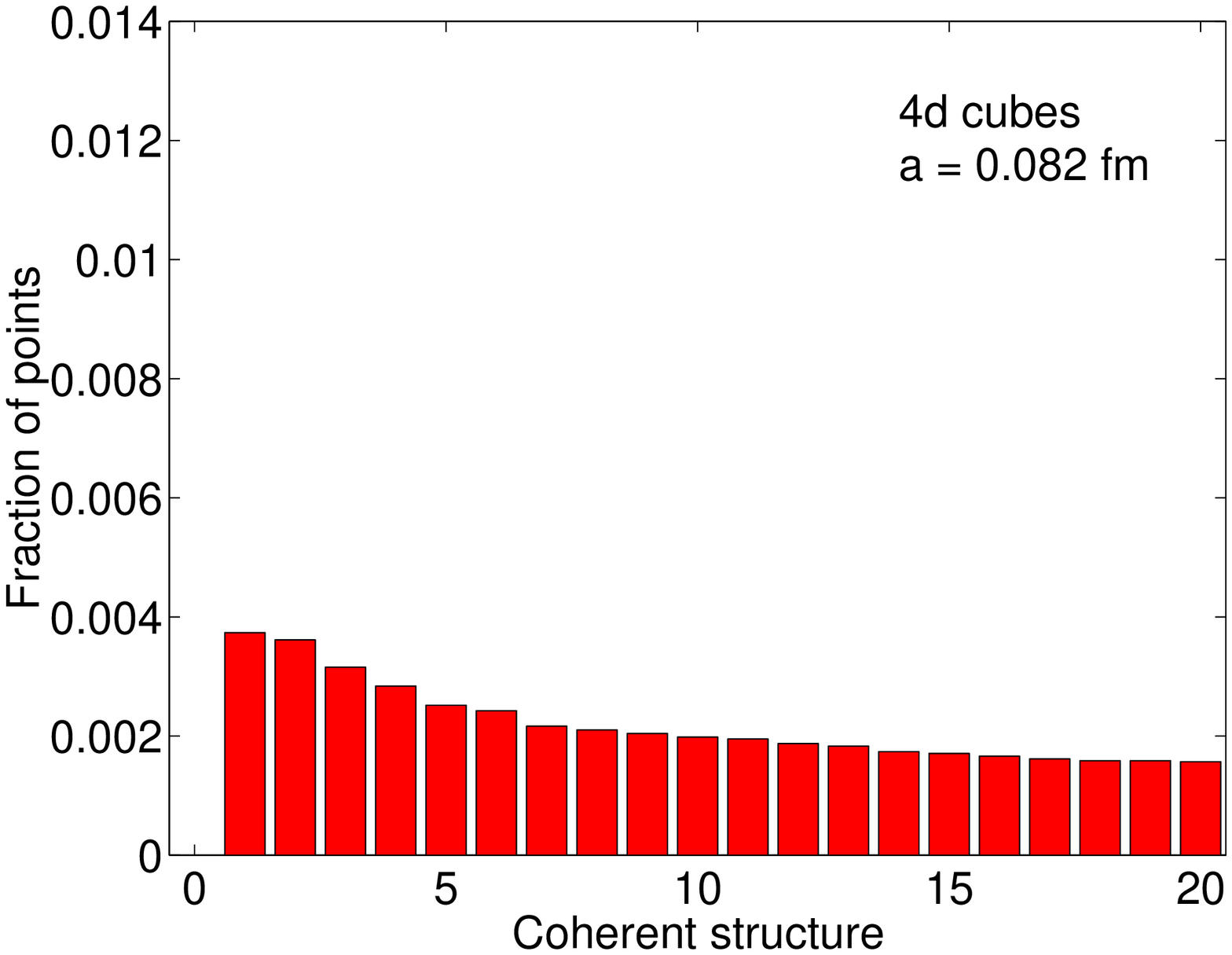}
\hspace*{-0.3in}
\includegraphics[height=4.65cm]{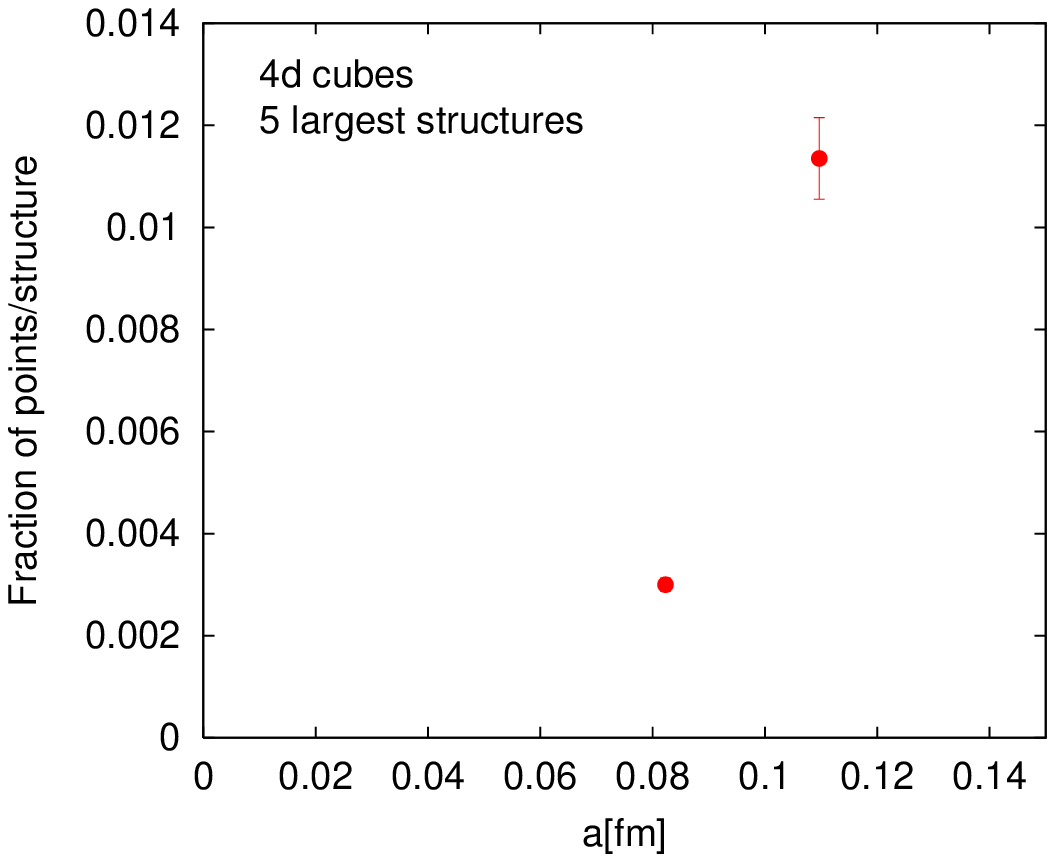}
\\
\hspace*{-0.1in}
\includegraphics[height=4.8cm]{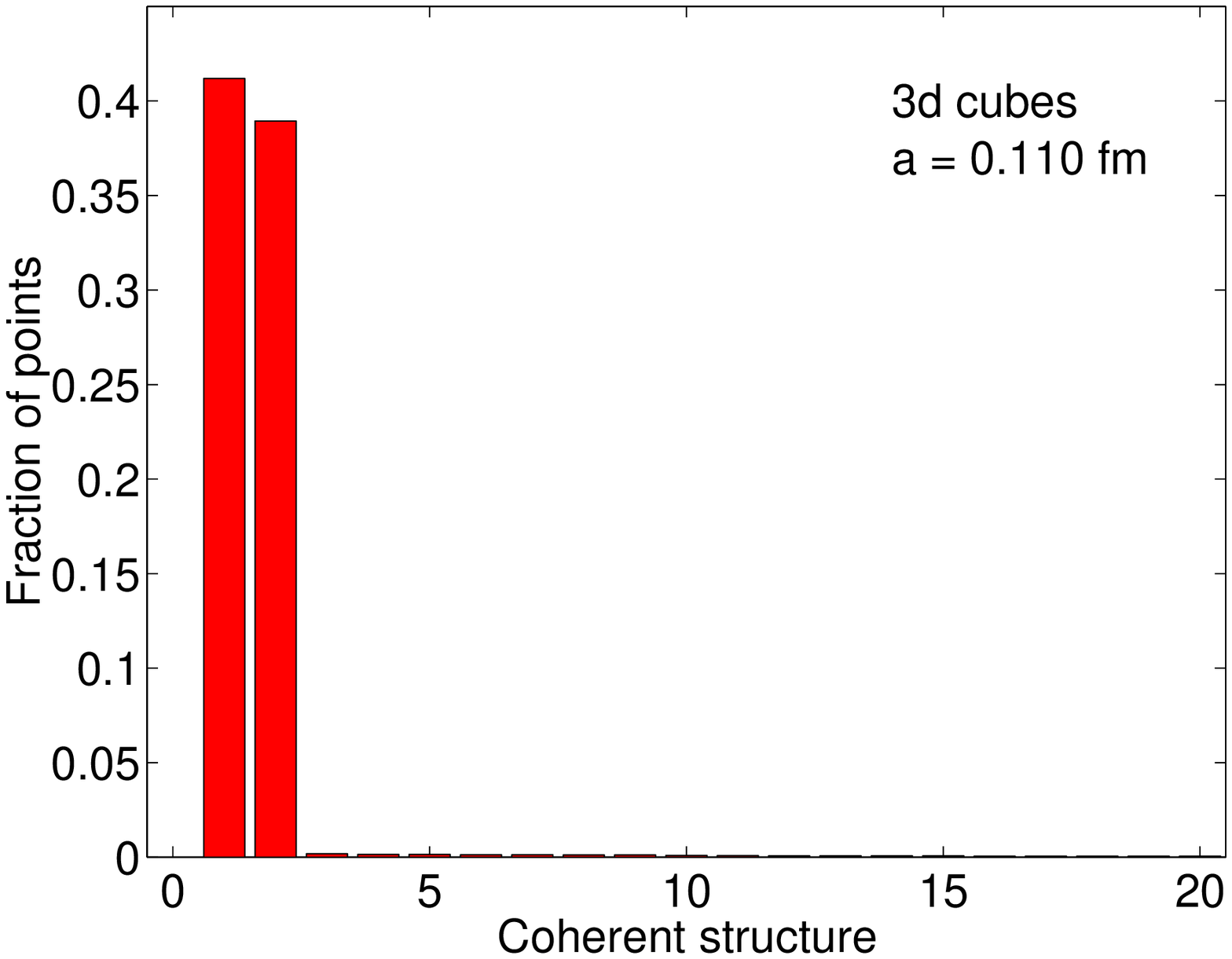}
\hspace*{-0.3in}
\includegraphics[height=4.8cm]{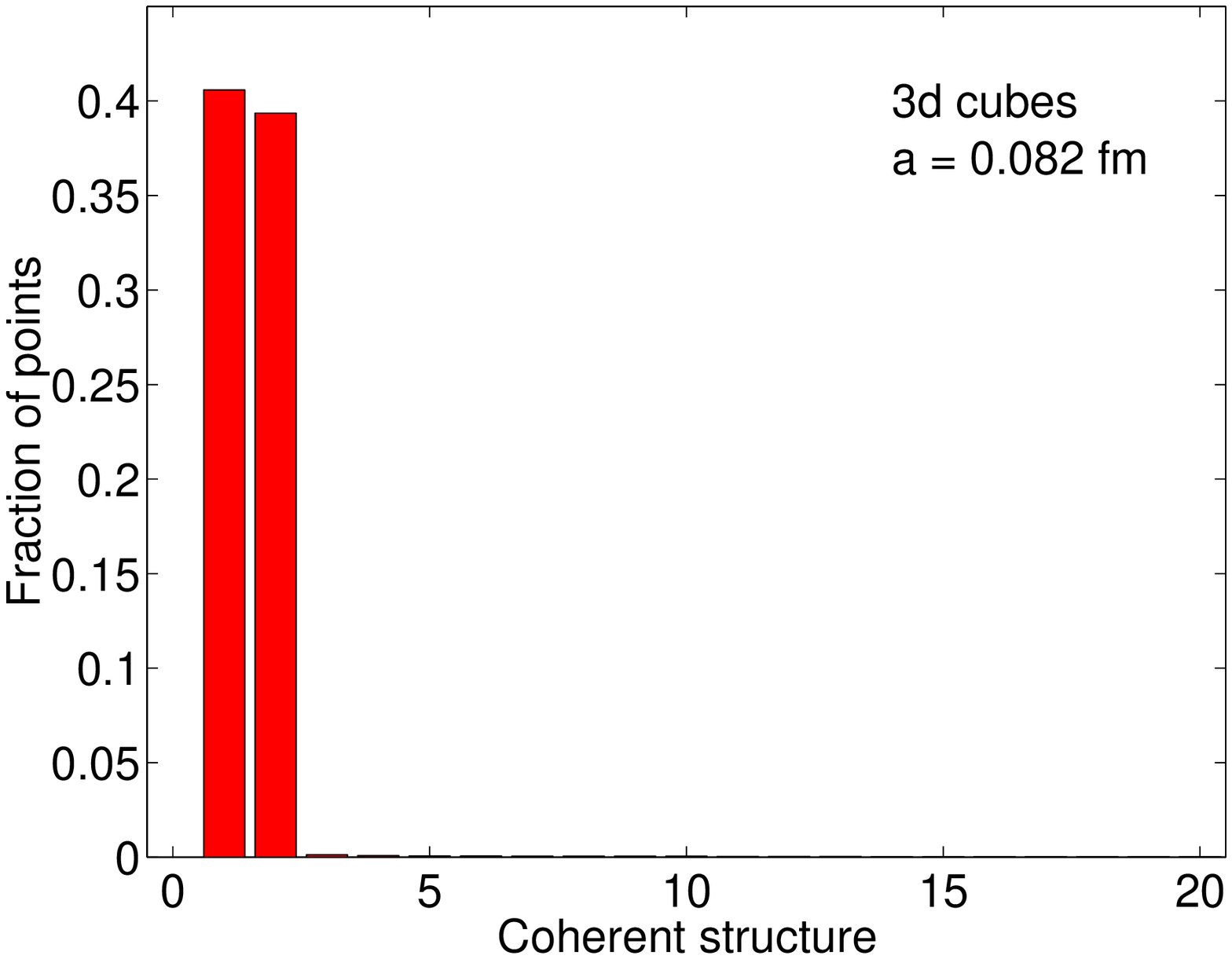}
\hspace*{-0.3in}
\includegraphics[height=4.65cm]{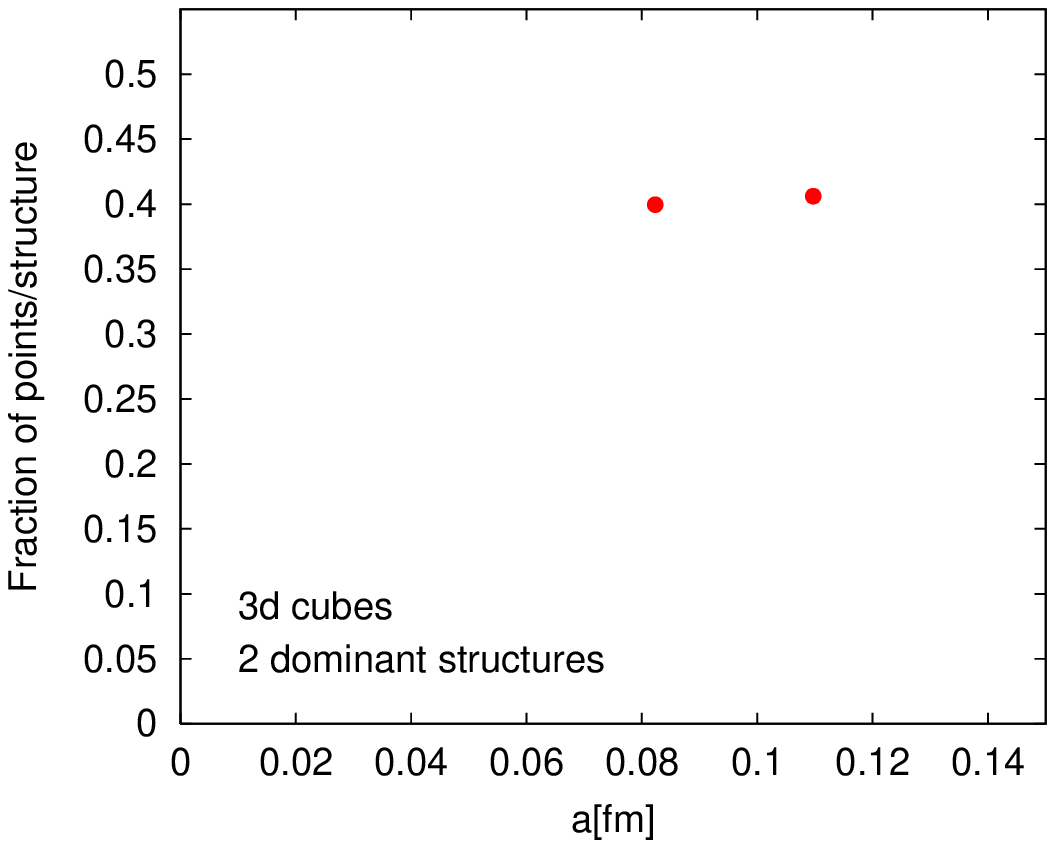}
\vspace*{-0.08in}
\caption{Top: The fraction of space-time points occupied by the largest 20 
structures built of 4-d cubes on a typical configuration from ensemble $\cE_1$ 
(left) and $\cE_2$ (middle). The average fraction from the 5 largest structures
is also shown (right). Bottom: The same as top but for structures built of 3-d 
cubes. There are error bars on the rhs plots.}
\vspace*{-0.08in}
\label{fig:05}
\end{figure*}

For the numerical study we use $q_x$ associated with the overlap Dirac operator~\cite{Neu98A}, 
whose suitability for this purpose was confirmed in Ref.~\cite{Hor02D}. To sample the QCD 
vacuum, we work with Wilson gauge backgrounds as summarized in Table~\ref{table:1}. 
Configurations are separated by 20,000 sweeps and the Sommer parameter serves to set the scale 
using the interpolation formula of Ref.~\cite{Sommer}. Since the computation is extremely 
time-consuming, the ensembles are small. However, the attempts to understand QCD vacuum
structure in the path integral framework implicitly rely on the expectation that relevant 
individual configurations exhibit common qualitative properties that are ``typical'' of the QCD 
vacuum, so that interesting results should be obtainable from even a few configurations. 
In our case, we found that for the aspects discussed below, the behavior is remarkably stable 
from configuration to configuration (both qualitatively and quantitatively). In fact, one 
could infer the main results presented here by studying just a single configuration from each 
ensemble. We are thus confident that our conclusions are not affected by low statistics. 
Some preliminary results related to this work were presented in Ref.~\cite{Hor02E}. Details 
of the fermion implementation can be found in Ref.~\cite{Hor02A}.

{\bf 1.} We start by verifying that $q_x$ is indeed not dominated by coherent 4-d 
structures, as argued above. Regardless of their shapes, the possible candidates 
for such structures must be built of elementary lattice 4-d cubes~\cite{Cubes} that are 
themselves coherent (their sites have the same sign of $q_x$). In fact, we can define the 
coherent regions on the lattice by finding all coherent 4-d cubes $\cC_x$ (labeled by origin $x$) 
and identifying maximal {\it connected} regions built from such cubes. We will say that 
lattice region $\cR =\{\,{\cal C}_{x_i}; i=1,\ldots,N\,\}$ is connected, if for arbitrary 
$\cC_{x_i},\cC_{x_j}\in\cR$ there is a sequence $\cC_{y_k}\,,k=1,\ldots,n\,,$ such that 
(a) $\cC_{y_k}\in \cR$, (b) $y_1=x_i$, $y_n=x_j$, and (c) $\cC_{y_k}$ and  
$\cC_{y_{k+1}}$ share a common face (3-d cube). 

We have determined all such regions ({\it ``structures''}) $\cR_k$ present in a given 
configuration, and ordered them by the number of lattice sites in the structure, $N_k$. 
If there are 4-d structures of finite physical size dominating in the continuum limit, then 
at least the largest of these coherent regions should exhibit scaling behavior when the lattice 
spacing is changed. Since the ensembles $\cE_1$, $\cE_2$ have the same physical volumes, 
the ratios $N_k/V$ ($V$ is the number of lattice points) for typical largest structures 
should not depend on the lattice spacing. In Fig.~\ref{fig:05} (top) we show these ratios 
for a configuration from both ensembles. Rather than being similar, the ratios drop 
dramatically at the smaller lattice spacing. To see this on average, we compute 
the mean value of the ratio including the 5 largest structures per configuration. 
Fig.~\ref{fig:05} reveals that the ratio not only drops, but it drops by a factor larger 
than the ratio of lattice volumes. This is confirmed by the fact that the average number 
of 4-d cubes per large structure is about $35.3$ at $a=0.110$, and is about $29.4$ at 
$a=0.082$. The volume of structures thus decreases even in lattice units. These 
considerations not only confirm that 4-d structures do not dominate topological charge 
fluctuations, but also suggest that such structures might not occur in the continuum at all. 
Indeed, in physical units, structures $\cR_k$ observed on the lattice appear to shrink 
to mere points in the continuum limit.

{\bf 2.} We now concentrate on the possibility that $q_x$ exhibits coherence on 
lower-dimensional subsets of 4-d space-time. As a first step, we search for maximal coherent 
regions built from elementary 3-d cubes connected through 2-d faces. Such regions define 
connected 3-d lattice hypersurfaces embedded in the 4-d space-time. We find that all of 
the configurations studied (see Table~\ref{table:1}) exhibit a remarkably similar 
coherent behavior. The typical situation 
for a configuration from ensembles $\cE_1$, $\cE_2$ is shown in Fig.~\ref{fig:05} (bottom). 
Contrary to the case of 4-d cubes, there are only two (oppositely charged) coherent 
regions dominating the behavior of $q_x$. Together, these folded lattice ``sheets'' cover 
about 80\% of sites and practically fill the space-time. This situation is insensitive 
to the change of the lattice spacing and our data indicate that it will survive 
the continuum limit (bottom right of Fig.~\ref{fig:05}). 

\begin{figure}[h]
\vspace*{-0.08in}
  \centerline{
     \includegraphics[width=4.5cm]{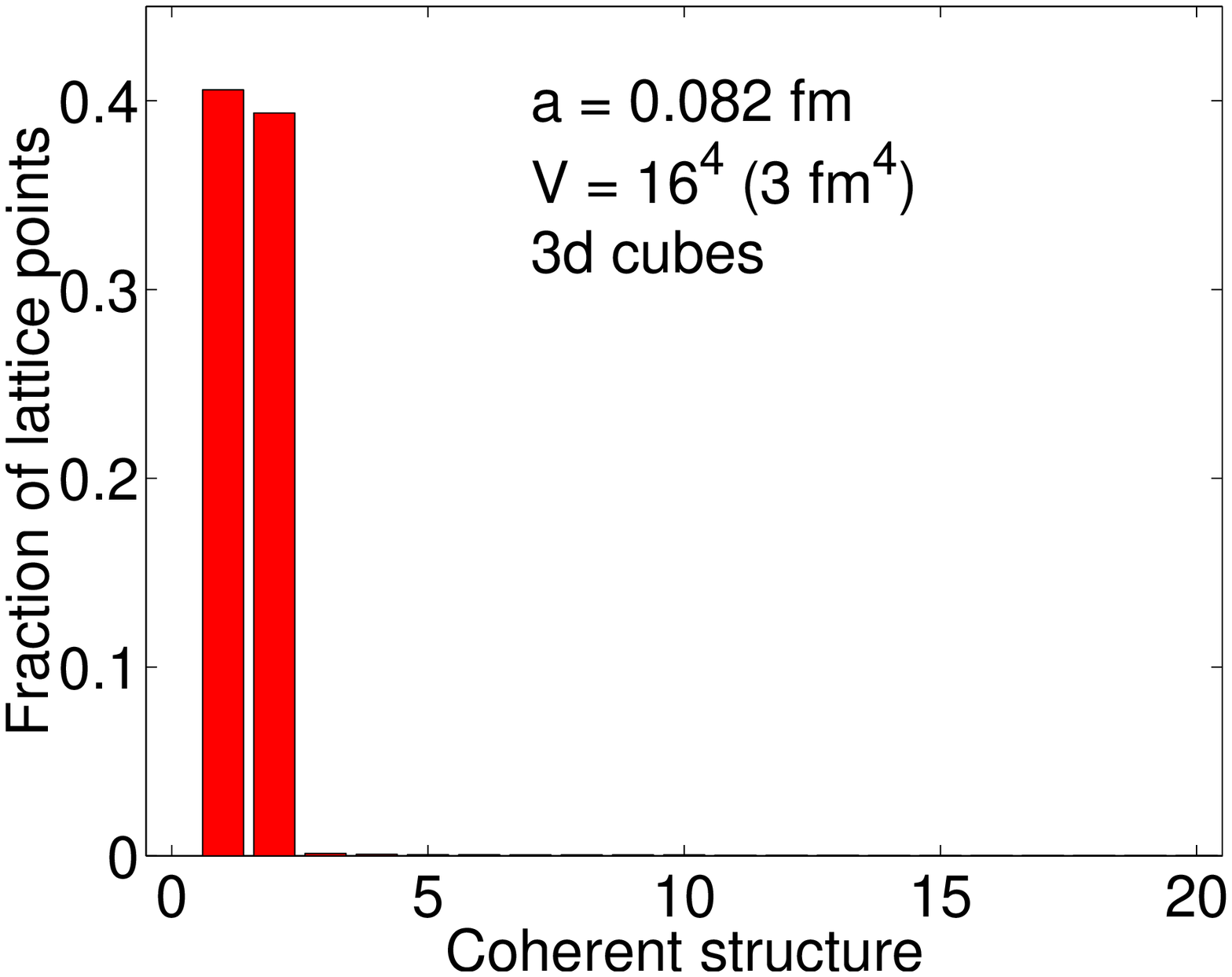}
      \hspace*{-0.15in}
     \includegraphics[width=4.5cm]{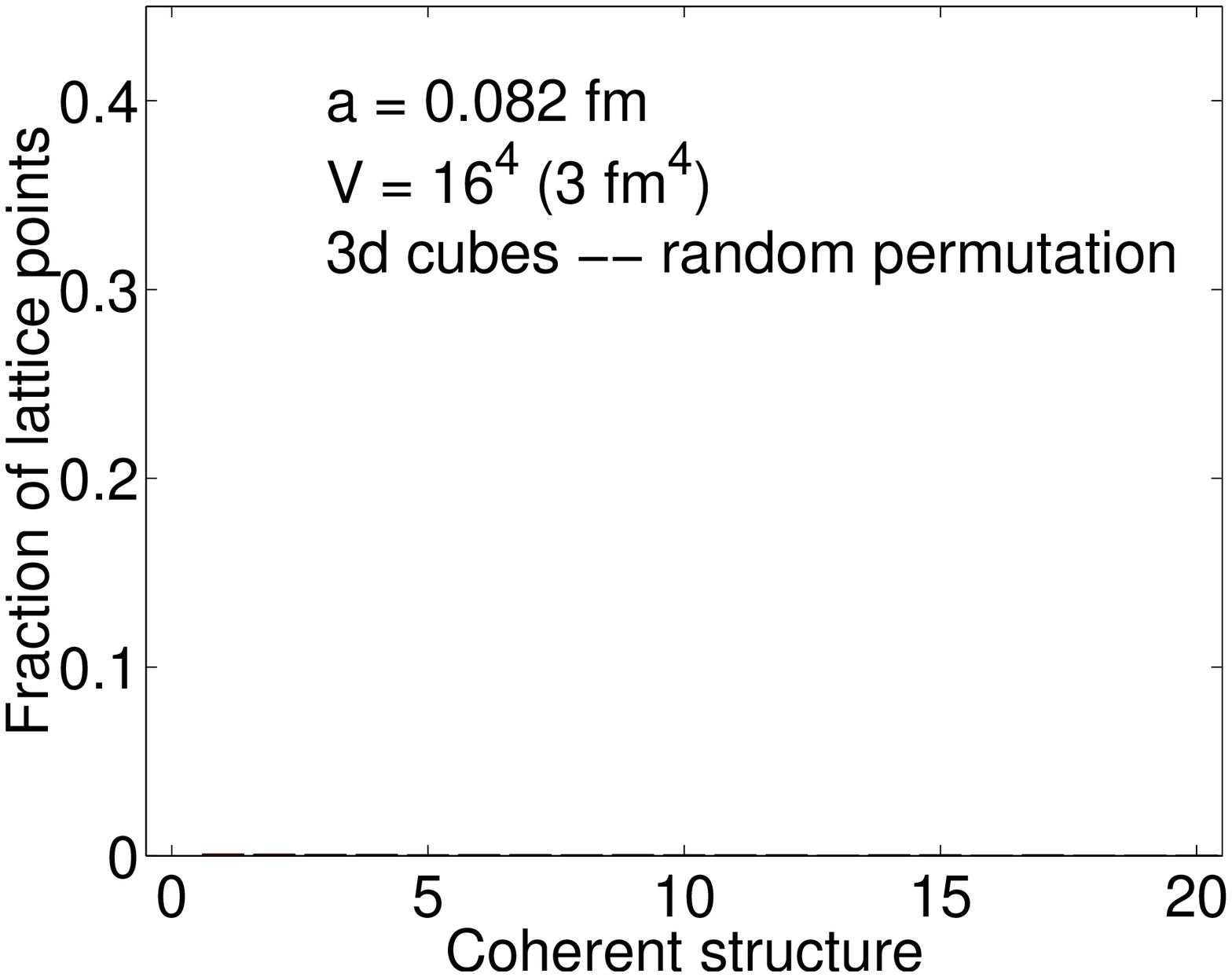}
   }
\vspace*{-0.04in}
\caption{A typical coherent structure before and after the random permutation of 
sites. On the common scale, the fractions on the r.h.s. plot are barely visible. 
Configuration from $\cE_2$.}
\vspace*{-0.0in}
\label{fig:10}
\end{figure}

We wish to emphasize two points. First, we have tested whether the existence of the 2-sheet 
structure reflects a specific order present in $q_x$, or if it could occur in a random 
situation as well. To do that, we have performed a random permutation of sites $p(x)$ 
on our lattices and looked for the coherent behavior in $q_{p(x)}$. The typical result 
of such a test is shown in Fig.~\ref{fig:10}. The 2-sheet structure disappears and 
the fractions involved are almost invisible on the scale of the original
plot. Secondly, the existence of (almost) space-filling 3-d {\it lattice} sheets is not 
sufficient to conclude that the local dimension of maximal coherent regions in the
continuum limit is three. The determination of this dimension requires a scaling analysis 
which is in preparation. While possibly less than three, the dimension is at least
one (see part {\bf 4} below). Note that this issue is not addressed 
by the lower right-hand plot of Fig.~\ref{fig:05}. There the implied scaling relates 
to the fact that the sheets occupy a macroscopic fraction of space-time and hence 
involve an infinitely large lower-dimensional manifold in the continuum limit 
(in the 4-d volume of finite physical size).

{\bf 3.} Since the 2-sheet structure almost fills the space-time, it is expected that 
for a generic point $x$ on a given sheet, there are points $y$ on the same sheet 
separated from $x$ by maximal space-time distances possible. Because the sheet 
is path-connected, such points can be connected by paths within the sheet. Thus, from 
the geometric point of view there is no specific long-distance scale (other than 
infrared cutoff) associated with global behavior of the sheet. We will refer to such 
a situation occuring in the coherent structure as the {\em super-long-distance} 
property~\cite{Percol}.

\begin{figure}[b]
\vspace*{-0.10in}
\hspace*{-0.15in}
\centerline{
\includegraphics[height=3.5cm]{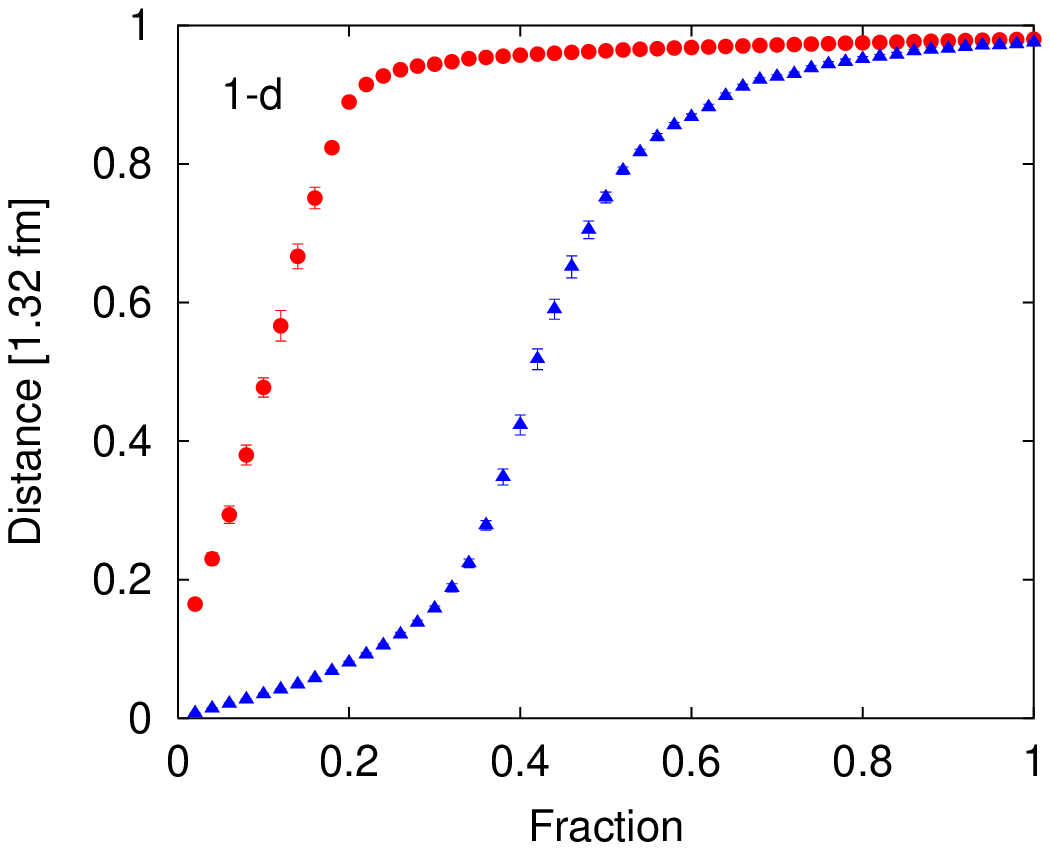}
\hspace*{-0.15in}
\includegraphics[height=3.5cm]{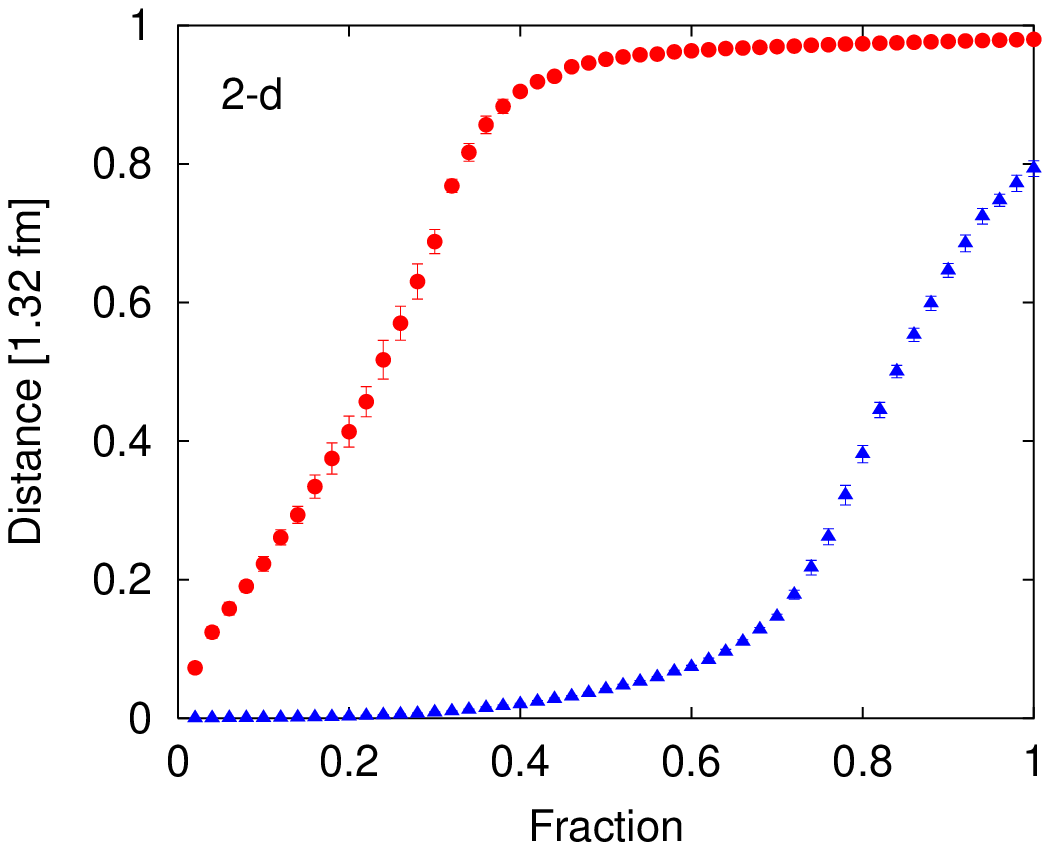}
}
\\
\hspace*{-0.07in}
\centerline{
\includegraphics[height=3.5cm]{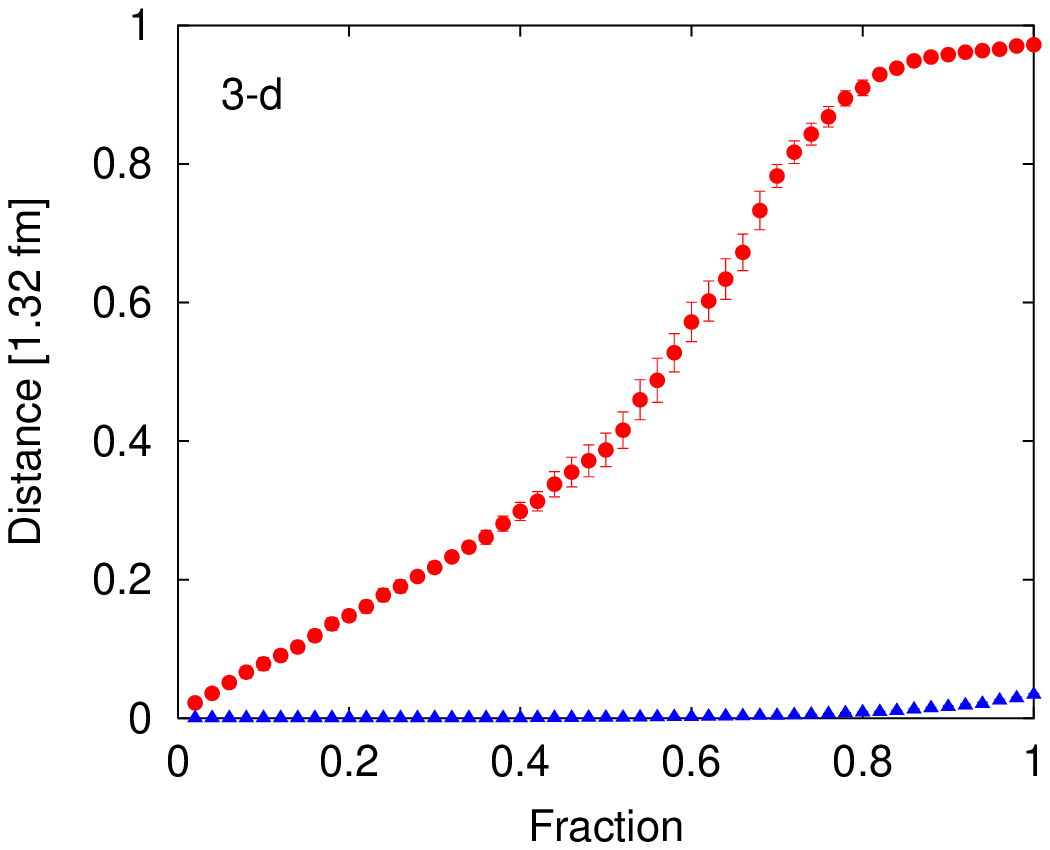}
\hspace*{-0.15in}
\includegraphics[height=3.5cm]{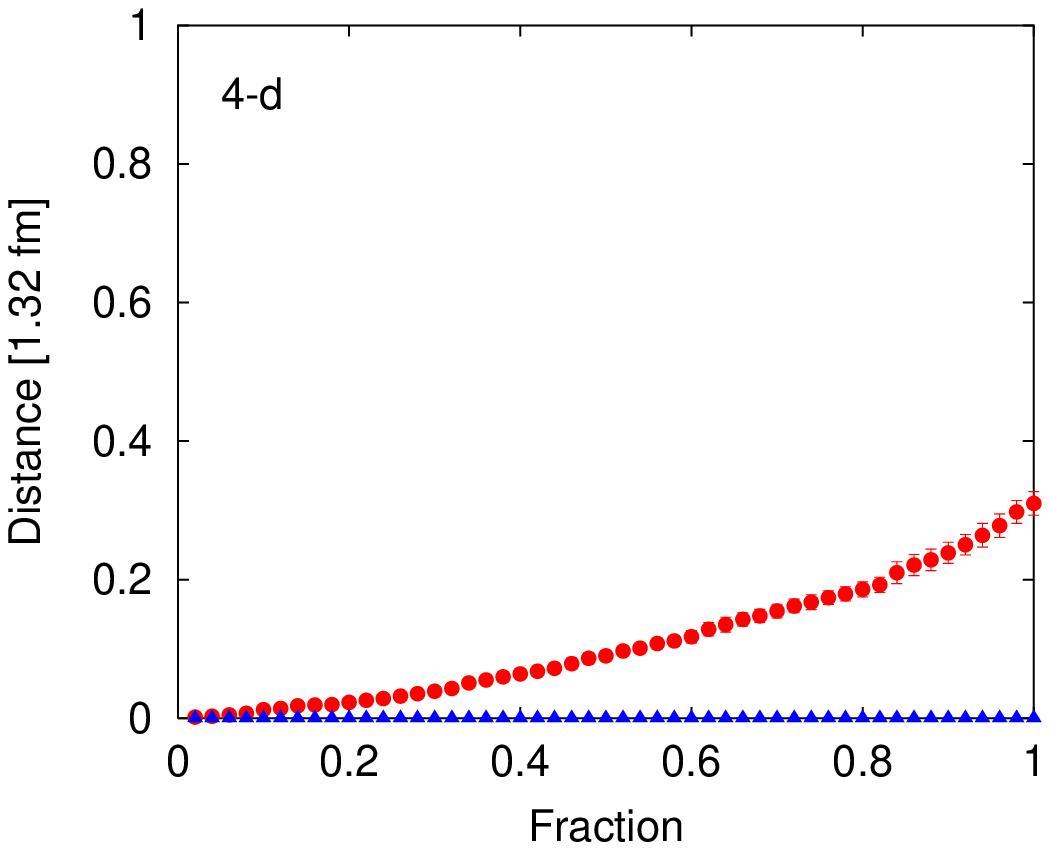}
}
\vspace*{-0.15in}
\caption{The $f$-dependence of $r^{0.01,f}_d$ for ensemble $\cE_1$ (upper curves of each plot) 
and after random permutation of sites (lower curves of each plot). Results using 1,2,3 and 
4-dimensional hypercubes are shown.}
\vspace*{-0.05in}
\label{fig:15}
\end{figure}

The super-long-distance property might be relevant to the physics of the QCD vacuum, 
e.g.\ to Goldstone boson propagation, if it remains true also if one concentrates 
on regions with largest topological charge density. Indeed, 
the physics of the vacuum is expected to be mostly driven by fluctuations with intense 
fields. For a given configuration, let $\cS^f$ ($0<f\le 1$) be the subset of the lattice 
containing the $f V$ most intense sites $x$, as ranked by $|q_x|$. Within $\cS^f$, there 
are maximal coherent regions $\cR^{d,f}_k\subset \cS^f$, $k=1,2,...,N^{d,f}$, built from 
$d$-dimensional cubes connected through $(d\!-\!1)$-dimensional faces. For $x\in\cS^f$ 
one can find a maximal Euclidean distance $r^f_{d,x}$ reachable from $x$ by traveling 
on a path within a single structure $\cR^{d,f}_k\ni x$.$\;$ Note that $r^f_{d,x}=0$ for 
points that do not belong to any structure. Considering a fixed small fraction $f_0<f$, 
we calculate the average of $r^f_{d,x}$ over $x\in \cS^{f_0}$, 
i.e. $r^{f_0,f}_d\equiv \langle r^f_{d,x}\rangle_{\cS^{f_0}}$. This provides us with 
the measure (scale) of how extended the coherent structure typically is. 
In Fig.~\ref{fig:15} we show the $f$-dependence of $r^{0.01,f}_d$ measured in units of 
the largest Euclidean distance $(\sum_{i=1}^4 L_i^2)^{1/2}/2$ ($L_i$ are the sizes of 
the lattice box). The average is taken over the ensemble $\cE_1$ and the observed behavior 
is typical (and stable) for $f_0$ of a few percent. Also shown is the result after performing 
the random permutation of sites for every configuration.

Note that in the case of 4-d cubes the structure never becomes super-long-distance. In fact, 
we expect that the maximal distance will approach zero in the continuum limit at arbitrary $f$. 
For d = 3, 2, 1, we observe a well-defined transition to a super-long-distance state which
is certainly present at $f\approx$ 0.8, 0.4, 0.2 respectively ($r^{0.01,f}_d\approx 0.9$ at 
those fractions). These values are insensitive to the change of the lattice spacing. We thus 
see that after removing the low-intensity points, the super-long-distance structure remains 
at the core if lower-dimensional elementary cubes are used. The minimal 
structure with this property is exposed with 1-d cubes (links) and we will refer to it 
as the {\it skeleton}. In our numerical experiments, the maximal space-time distances 
start occuring already at about $f=0.16$ while the average saturates at about $f=0.20$. 
We thus take $f=0.18$ as our initial reference value for the skeleton. As further 
evidence that the observed structure carries a significant amount of order, we note that 
at $f=0.18$ (with the skeleton already formed) one is still confined to average maximal 
distances of a single lattice spacing when random permutation of sites is performed.

\begin{figure}[t]
\vspace*{-0.06in}
  \centerline{
     \includegraphics[width=7.0cm]{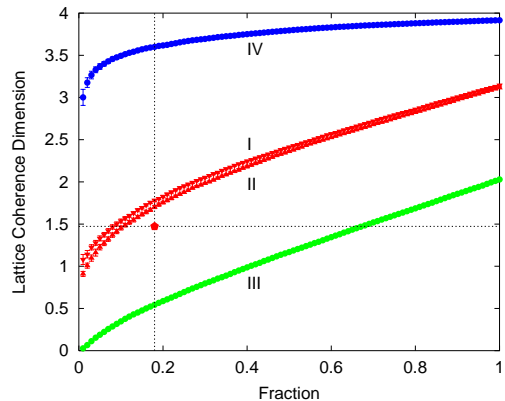}
             }
\vspace*{-0.05in}
\caption{Lattice coherence dimension. Explanation in text.}
\vspace*{-0.17in}
\label{fig:20}
\end{figure}

{\bf 4.} Even without an extensive scaling analysis, it is possible to obtain useful 
insight into the question of the actual space-time dimension involved in 
the skeleton. More precisely, we will discuss the {\it local} dimension. 
To appreciate the difference, consider e.g. a network built from 1-d lines, but organized 
globally into a 2-d structure (``mesh'') with a typical physical distance $l$ among 
the nodes. Locally, such a manifold is 1-dimensional, but on the scales larger than $l$ it 
will appear 2-dimensional. We will consider the global structure from this point of view 
elsewhere.

Our tool here will be the {\it lattice local dimension} $d^{L}(\cS)$ which we
define for an arbitrary subset $\cS$ of the lattice. For $x\in \cS$ let $d_x^L$ be the maximal 
dimension $d$ of the elementary cube $\cC^d_y$ such that $x\in\cC^d_y$ and 
$\cC^d_y\subset \cS$, i.e. the maximal dimension of the cube within $\cS$ containing $x$. 
Then $d^L(\cS) \equiv \langle d_x^L \rangle_\cS$ is the average of this local dimension
over $\cS$. While $d^L(\cS)$ is strictly a lattice notion, it can still be quite useful. 
In particular, if there is an underlying set $\cS^c$ in the continuum characterized 
by a single local dimension $d(\cS^c)<4$, then the continuum limit of $d^L(\cS)$ typically 
provides an upper bound for $d(\cS^c)$. For example, consider a line of non-zero physical 
length in the continuum. At sufficiently small lattice spacing this might appear as 
the connected structure built predominantly of bonds, in which case the continuum limit 
of $d^L(\cS)$ coincides with $d(\cS^c)$. However, it might also come as a string of mainly 
4-d cubes at arbitrarily small lattice spacing, and then the limit of $d^L(\cS)$ will 
be significantly larger than $d(\cS^c)$. At the same time, $d^L(\cS)\ge 1$ for 
the {\it connected} lattice structures and the physical dimension will not be underestimated.

Since the super-long-distance property involves connected parts spreading over maximal distances, 
and since it is insensitive to the change of lattice spacing, we can conclude that the skeleton 
is at least 1-dimensional. To see whether higher dimensions (less than four) could be relevant 
we consider 
sets $\cS^f$ and their partitions $\cS^f=\cS^{f+}\cup\cS^{f-}$ into coherent subsets containing 
sites with positive and negative sign of $q_x$. In Fig.~\ref{fig:20} we show the $f$-dependence 
of the average {\it lattice coherence dimension} 
$\langle d^L(\cS^{f\pm})\rangle$ for ensembles $\cE_1$ (I) and $\cE_2$ (II). 
We see that the dimension slightly {\it decreases} as we approach the continuum limit. 
The naive linear extrapolation for the skeleton value f = 0.18 (pentagon) gives 
$\langle d^L(\cS^{f\pm})\rangle\approx 1.5$ in the continuum limit and we thus conclude that 
the physical local dimension associated with the skeleton is 
$1\le d \:\raisebox{-0.8ex}{$\stackrel{\raisebox{-1.5ex}{$\textstyle<$}}{\sim}$}\:1.5$. 
Assuming that the local structure is not fractal (an option to be investigated) 
we are thus led to the conclusion that the skeleton is locally 1-dimensional. 

For comparison, we show in Fig.~\ref{fig:20} also $\langle d^L(\cS^{f\pm})\rangle$ for $\cE_2$ 
after random permutation of sites (III), and the result for effective density 
$q_x^{(\Lambda = {\rm 750\, MeV})}$ at the same lattice spacing (IV). As expected, the curve 
(III) lies lower and tends to zero smoothly at small fractions. On the other hand, for
the effective density one naively expects 4-d coherence on the scale $1/\Lambda$ in the continuum 
limit, and thus the lattice coherence dimension is much higher.


\begin{figure}[t]
 \vspace*{-1.30in}
 \centerline{
 \includegraphics[width=9.4cm]{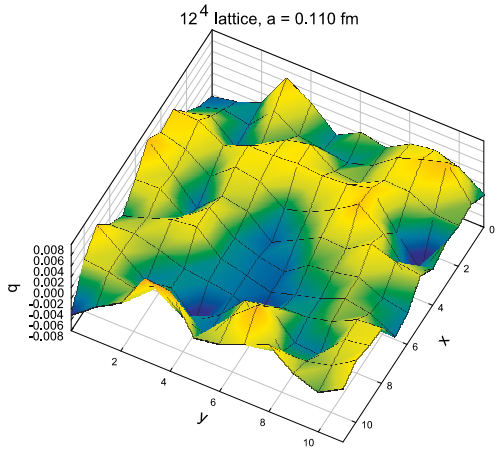}
            }
\vspace*{-1.05in}
\caption{Generic behavior of $q_x$ on a 2-d section of space-time (configuration from $\cE_1$).}
\vspace*{-0.15in}
\end{figure}

{\bf 5.} Finally, to start developing a visual intuition for the kind of structure we are
observing, we now proceed to graphically verify the existence of the coherent sheets described 
in part {\bf 2}. To do that, we have studied the graphs of $q_x$ on 2-d sections of the underlying
space-time torus. The {\em generic} behavior of $q_x$ is shown in Fig.~5 (configuration 
from ensemble $\cE_1$). One can clearly observe the enhancement of the structure with linear 
``ridges'' spreading over the whole system. It is easy to see that this is precisely the behavior 
expected from the (almost) space-filling super-long-distance sheets. Indeed, recall that if 
$\Omega$ is an arbitrary base manifold and $\Omega_1,\Omega_2 \subset \Omega$, then 
$\dim ( \Omega_1 \cap \Omega_2 ) = \dim (\Omega_1) + \dim(\Omega_2) -  \dim(\Omega)$
generically. If we identify $\Omega$ with the 4-d space-time torus, $\Omega_1$ with the 2-d 
torus of the section plane, and $\Omega_2$  with the manifold of the structure, we get
\begin{equation}
   \dim ( \Omega_1 \cap \Omega_2 ) \,=\, \dim(\Omega_2) \,- \, 2
   \label{eq:30}
\end{equation}
Thus if $\Omega_2$ is a 3-d super-long-distance hypersurface, we should generically see very 
extended 1-d regions of sign-coherence, which is precisely what is observed. Thus, at the lattice 
cutoff of about 2 GeV the sheets behave as 3-d hypersurfaces. The precise local dimension 
of maximal coherent regions in the continuum limit will be studied elsewhere.

In this work we have addressed the following question. Can we identify a fundamental (incorporating
all scales) ordered structure in topological charge density for typical configurations contributing 
to the pure-glue QCD path integral? In other words, using nothing more than a {\em local} operator 
$q_x$ to calculate space-time distribution of topological charge, can one detect well-defined 
patterns in these distributions? While such structure has not been observed before, we have argued 
that the special properties of the newly available topological charge densities associated with 
Ginsparg-Wilson fermions give reasons to re-examine this question in detail. In particular, 
the suppression of ultraviolet noise by {\it chiral smoothing} suggests that the problem of 
ultraviolet dominance might be solved and the underlying order could be revealed. Performing 
numerical experiments, we have then provided evidence for a fundamental structure of this type
by showing that topological charge exhibits long-range order (sign-coherence) on 
lower-dimensional subsets of the 4-d Euclidean space. In this initial study we have concentrated 
on certain geometrical properties of this structure. Indeed, if topological charge fluctuations 
play a role in important aspects of the QCD vacuum (such as S$\chi$SB), then some geometrical 
features should reflect that. While it is not clear yet which structural aspects are physically 
most relevant, we emphasize two that we find most intriguing: (i) Space-time regions with intense 
topological charge density are 
organized into a maximally extended (super-long-distance) structure ({\em ``skeleton''}). This 
suggests that there is no specific long-distance scale associated with {\em individual} geometrical 
objects. Rather, there is a global low-dimensional structure that might have to be considered 
as a whole. (Distance scales associated with $\Lambda_{QCD}$ could be reflected by local features 
of such global structure.) We speculate that the super-long-distance character of the skeleton 
might be naturally associated with the long-range propagation of Goldstone pions. It will be 
demonstrated elsewhere that the super-long-distance property is inherited and {\it enhanced} 
at low energy. (ii) Our data indicates that the skeleton is a locally-linear folded structure, much 
like a protein or a neural network in 3-d. One aspect that might be significant about this result 
is that the propagation of massless quarks on a 1-d manifold naturally leads to a non-zero density 
of Dirac eigenvalues around zero, and hence S$\chi$SB (see e.g. Ref.~\cite{Tik87} for discussion
of similar issues). The possibility that such a scenario might be relevant for QCD is being studied.



\begin{thebibliography}{99}
\bibitem{Cubes} We will refer to elementary ``links, squares, cubes and hypercubes'' 
   of 4-dimensional lattice uniformly as 1-d cubes, 2-d cubes, etc.
\bibitem{Percol} It is tempting to refer to such structure also as a ``percolating
   structure''. However, at this stage it is not clear whether similarities to
   percolation go beyond the unbounded nature of the structure.
\end{thebibliography}
\end{document}